\documentstyle[12pt]{article}

\begin{document}

\title{ Radiation pressure approach to the repulsive Casimir force\footnote{
Revised summary of talk given to the ITAMP Topical Group on {\it
CasimirForces},
Cambridge, MA, March 16-27, 1998.}\\} 
\author{V. Hushwater \\ 
{\em Physics Department, University of Maryland, College Park,} \\ 
{\em MD 20742}}
\date{June 9, 1999}

\maketitle

\vspace{2mm} Radiation pressure\\ 
--------------------------- \\
We study the Casimir force
between 
a perfectly conducting and an
infinitely permeable plate with the radiation pressure approach. This
method illustrates how a repulsive force arises as a consequence of the
redistribution of the vacuum-field modes corresponding to specific
boundary conditions. We discuss also how the method of the zero-point
radiation pressure follows from QED.\\ 
---------------------------
\vspace*{5mm}

\pagebreak

\begin{quote}
\em `Understanding of signs is a sign of understanding.' J. 
Sucher [1]
\end{quote}

At the end of his work on a force between 
two perfectly conducting parallel plates in vacuum as a consequence
of the change in zero-point energy [2], Casimir proposed that 
this force can be interpreted as radiation pressure from the 
vacuum field. Later this interpretation was reintroduced by several other 
authors [3 - 5] and used [4, 5] for calculation of the Casimir force
in such a case. As was mentioned in [5],   
the  radiation pressure approach can be
systematically developed on the basis of QED. Actually this was
 already shown in 1969 by L. Brown and J. Maclay [6] (who also
used this
approach
to compute the attractive Casimir force). We will discuss this question
in some detail in Appendix.      

P. Milonni, R. Cook and M. Goggin [5] noticed a puzling character of the
vacuum radiation pressure:  In the case of two perfectly conducting 
plates the modes of the 
vacuum field in the space outside the plates form a continuum, corresponding 
to arbitrary wave vector $\bf k \rm$ , whereas those inside are 
restricted to discrete 
values of $k_{z}$  (with the $z$-axis perpendicular to the plates). 
So there are  
"more"\footnote{Quotation marks since we compare infinite numbers!} modes
outside to
push the plates together by the radiation pressure 
than there are modes between the plates to push them apart. This results
in  the  attractive Casimir force\footnote{This 
argument is also given in [7] 
and [8].}.  However,  they 
concluded that this argument is superficial, since it cannot explain a radially
outward Casimir force, known in the case of a  spherically  conducting shell, 
in spite of the fact that there also should be  "fewer" modes of the 
vacuum field inside the shell than there are outside.

In order to resolve the puzzle we should look more carefully to how
boundaries 
affect the zero-point field. Doing this one can
 notice that, loosely speaking, as a whole there are no fewer normal  
modes inside confined volumes than there are in in free space; 
they  just  
are shifted to other frequencies. As it turns out this question was
discussed in a rigorous way by G. Barton [9 - 10] and mentioned in [11]. 
Not knowing his results I came to the same conclusion from the following
considerations. Consider, e. g., the
space
between two perfectly conducting plates separated by the distance $ l$ .
 As was mentioned above, the vacuum 
electromagnetic
 field is here a sum of modes (standing waves) permitted  by the boundary  
conditions. 
Each such a mode is mathematically equivalent to a harmonic oscillator of the 
same frequency.  When the distance between plates is very large we 
actually have the free space case. If now one adiabatically moves  the
walls  
toward 
each other, the "electromagnetic oscillators" will remain in the ground state; 
only their frequencies will be shifted to values corresponding to the changed 
boundary conditions [12]. Thus, the main effect of the boundaries is to 
redistribute normal modes : for some frequencies $\omega $ there are more 
modes than in free space, for others there are fewer.

One can check this conclusion by comparison of the mode spectral
densities. In free space the mode spectral density $\rho (\omega)$  
in a volume V is 
\begin{equation}
                  \rho _{0}(\omega ) = \frac{V\omega^{2}}{\pi^{2} c^{3}}, 
\end{equation}
while inside the cavity between two perfectly conducting plates [13], 
where \break  $k_{z} = n\pi /l$, $ n = 1,2,...$,  
\begin{equation}
\rho_{1}(\omega ) = \rho_{0}(\omega )\frac{\omega _{0}}{\omega }[1/2 + 
\sum_{m=1}^{\infty} \theta ( \frac{\omega }{\omega _{0}}  - m)]. 
\end{equation}
Here $c$ is velosity of light, $\omega _{0} = c\pi /l$,  and $\theta $ is
the Heaviside step function. 

In the case of a perfectly conducting plate parallel to 
an infinitely permeable  plate, with the same separation $ l$,
$k_{z} = (n + 1/2)\pi /l$, $ n = 0,1,2,...$ [14] and therefore the mode 
spectral density is
\begin{equation}
\rho_{2}(\omega ) = \rho_0(\omega )\frac{\omega_0}{\omega} 
\sum_{m=0}^{\infty} \theta ( \frac{\omega }{\omega _{0}}  - m + 1/2).
\end{equation}
One can see that for about a half of the frequencies 
$\rho_{1}(\omega )$  
and $\rho_{2}(\omega )$ are greater than $\rho_{0}(\omega )$.

     It is just a result of the redistribution of normal modes  that  the 
pressure from the inside vacuum field $P_{out}$ is different than the 
oppositely directed pressure from a free space vacuum field $ P_{in}$  . As 
noted  above,  for two conducting plates $P_{in} > P_{out}$. However, since the 
redistribution of modes depends on  boundary conditions,  the relation 
between two pressures in other cases can be opposite.

As an example of such a case
 let us consider a perfectly conducting plate parallel to 
an infinitely permeable  plate, mentioned above.
If a plane wave has an angle of incidence $\theta $ the radiation pressure
exerted by such a wave on a plane, $P = 2w \cos ^{2}(\theta )$, where $w$ 
is the energy density. So a vacuum field mode of frequency $\omega $, 
which has an angle of incidence $\theta $, 
makes a contribution to the pressure 
\begin{equation}
P(\omega ) = 2 \frac{1}{2} \frac{h\omega }{2V} \cos ^{2}(\theta) =  
\frac{h\omega }{2V} (\frac{k_{z}}{k})^{2},  
\end{equation}
where $k = wc$ and $V$ is a quantization volume. A factor $1/2$ has been 
inserted because the zero-point energy of each mode is divided equally 
between waves propagating toward or away from each plate [5].

 Therefore we find for the net pressure $P = P_{out} - P_{in}$,
 where $P_{out}$ and $P_{in}$ are vacuum radiation pressures directed outward
and inward, correspondingly,
the expression    
\begin{eqnarray}
P &=& (\hbar c/(\pi^{2} l)^{2} \sum_{n = 0}^{\infty }
\int_{0}^{\infty} dk_{x} \int_{0}^{\infty} dk_{y} 
\frac{[(n + 1/2)\pi /l ]^{2}}{(k^{2}_{x} + k^{2}_{y} + [(n + 1/2)\pi
/l]^{2})
^{\frac{1}{2}}}  \nonumber \\
& & - (\hbar c/{\pi }^{3} ) \int_{0}^{\infty} dk_{x} \int_{0}^{\infty}
 dk_{y}\int_{0}^{\infty} dk_{z}\frac{ k^{2}_{z}}{(k^{2}_{x} + k^{2}_{y} 
+ k^{2}_{z})^{\frac{1}{2}}} ,    
\end{eqnarray}
which can be transformed  using  variables $ s \equiv
(l/\pi)^{2}(k^{2}_{x} +
 k^{2}_{y})$
  and $u \equiv (l/\pi )k_{z}$   into the form
\begin{eqnarray}
P & = & (\hbar c\pi/4 l^{4})[ \sum_{n = 0}^{\infty } (n + 
1/2)^{2} \int_{0}^{\infty} \frac{ds}
{(s  + (n + 1/2)^{2})^{\frac{1}{2}}}  \nonumber \\
& & - \int_{0}^{\infty} du u^{2}\int_{0}^{\infty}
\frac{ds}{(s  + u^{2})^{\frac{1}{2}}}] .  
\end{eqnarray}

 In order to regularize the divergent integrals in (6) a cutoff 
function \linebreak  
\mbox{$f_{a} ([s + u^{2}]^{1/2})$}   
must be introduced, with $f_{a} 
\rightarrow 1 $ when the parameter $a$ tends to,  say, zero:
\begin{equation}
         \lim_{a \rightarrow 0} f_{a}(p) \rightarrow 1 ,  
\end{equation}
here $p \equiv ([s + u^{2}]^{1/2})$.
Further, one requires that $f_{a} (p)$ vanish rapidly enough for $y 
\rightarrow \infty $ , 
\begin{equation}
          \lim_{p \rightarrow \infty} f_{a}(p) \rightarrow 0 . 
\end{equation}
so that the function           
\begin{equation}
               G_{a}(u) \equiv   u^{2}\int_{0}^{\infty}
 ds \frac{f_{a} ([s + u^{2}]^{\frac{1}{2}}}{(s  + u^{2})^{\frac{1}{2}}}  
\end{equation}
is finite and  $ G_{a}(\infty) = 0 $.  We can thus rewrite (6) as
\begin{equation}
P = lim_{a \rightarrow 0} (\hbar c\pi/4 l^{4})[ \sum_{n = 0}^{
\infty}
 G_{a}(n + 1/2) - \int_{0}^{\infty} du G_{a}(u)] . 
\end{equation}

Using the Euler-Maclaurin formula [15] one can find [12] that     
 in the limit  $a \rightarrow 
0 $  the difference  in  pressure  is  finite, independent of cutoff, and
reduces to  
\begin{equation}
P = (7/8)\hbar c\pi/240 l^{4} .
\end{equation}
This coincides with the expression for the Casimir force for  such  a  system 
obtained by Boyer [14], who used the energy  difference  method  and  a  
special (exponential) form of cutoff function.

Let us discuss shortly as to why the net vacuum radiation pressure
has positive sign in the case under consideration.
Let  $ H_{a} (u) \equiv G_{a} (n + 1/2)$  for 
$ n < u < n + 1$ , where $ n$  is  an integer.
It follows from (10) that the net sign of the vacuum radiation pressure 
depends on whether the area under the  step function $ H_{a} (u)$ is 
larger or smaller than  the area under $G_{a} (u)$.

It is not difficult to see that if  the  curvature  of $ G_{a} (u)$  is  
zero  for $n < u < n + 1$\nolinebreak[4] ,  the  area  difference between
$ H_{a} (u)$ and
 $ G_{a} (u)$ in  this  region  is equal  to  zero.
If the curvature is not zero the area difference will be bigger  or  smaller 
than these values depending on the sign of $ d^{2}G_{a}/du^{2}$ , i. e. on 
whether in the considered region $G_{a} (u)$ is concave or convex. 
 
One can check that $G_{a} (u)$ is primarily convex for any
acceptable cutoff function $f_{a}$. So the 
net area beneath $H_{a} (u)$ is  
larger than that beneath $G_{a} (u)$ and, therefore, in the case of one
 conducting 
and one permeable plate $P_{out} - P_{in}  > 0$  and we have repulsion.

    So we showed in a very simple case  how  zero-point  radiation 
pressure can lead to a repulsive rather than attractive Casimir force. It is 
the precise distribution of normal mode frequencies associated with specific 
boundary conditions, together with the fact that the function $F_{a} (u)$
is primarily convex, that determines the sign of the force. 
\vspace{5 mm}

{\Large\bf Appendix}\linebreak

Let us show how the method of the zero-point radiation pressure
follows from QED and that it is equivalent to the method of the change in 
the energy of quantum fluctuations of the electromagnetic field.

It follows from the operator Maxwell's equations that $\bf g$, the 
operator for the momentum density of the electromagnetic field  is 
\begin{equation}
{\bf g} = \frac{1}{8\pi}({\bf E}\times {\bf B} - {\bf B}\times {\bf E}),
\end{equation}
where ${\bf E} = {\bf E}({\bf r}, t)$ and ${\bf B}  = {\bf B}({\bf r}, t)$
are the Heisenberg operators of the electric and
 magnetic fields. $\bf g$ obeys the equation of continuity

\begin{equation}
\frac{\partial g^i}{\partial t} + \frac{\partial T^{ij}}{\partial x^{j}}
= 0,
\end{equation}
where $T^{ij}$, $i,j = 1, 2, 3$ is the spatial part of the $4$-dim.
energy-momentum
tensor operator $T^{\mu \lambda}$, $\mu , \lambda = 0, 1, 2, 3$ of 
the electromagnetic field$, x^{0} = ct, \linebreak x^{1} = x,
x^{2} = y, x^{3} = z$.
$T^{\mu \lambda}$ is a  certain combination of  electric and magnetic
fields components and, therefore, is an operator function of
$x^{\mu}$, the
form of which depends on the problem. 
We will consider below $T^{\mu \lambda}$ renormalized in such a way, that
its
expectation value  in the ground state of the electromagnetic field, 
$ <T^{\mu \lambda}>$, is zero in free space.\footnote{The 
following is basically a slightly modified derivation given
in [6].} 

Eq.(13)  describes the local conservation of the field momentum. 
Integrating  $\partial g^i /\partial t$
over a volume $V$ and using the divergence theorem one can find the force 
 exerted on this volume 
from the internal
electromagnetic field. The expectation value for a
force component, $F^i$ is
\begin{equation}
F^i = \int dv <\frac{\partial g^i}{\partial t}>
 = - \int dv <\frac{\partial T^{ij}}{\partial x^{j}}>
 = \int_{\Sigma}<T^{ij}>n^{i} da ,
\end{equation}
where $\bf n \rm$ is a unit outward normal to the
surface which surrounds volume 
$V$ and $da$ is an element of the surface area.  

Let us show that $F^i$ determined by (14) coincides with the force 
determined by the zero-point energy method:
\begin{equation}
 F^{i} = -\frac{\delta \varepsilon }{\delta x^{i}},
\end{equation}
where  $\delta \varepsilon $ is a infinitesimal change in the zero-point
energy of the electromagnetic field in 
the volume $V$ under the influence of a virtual
displacement  $\delta x^{i}$ of its surface.

Consider for 
simplicity the case of two  parallel infinite plates 
separated by a distance $l$, with the $z$-axis perpendicular to 
them. Each plate is either a perfect conductor or is infinitely
permeable. 
As follows from dimensional considerations, the 
energy per unite area of the electromagnetic field between the plates,
$\varepsilon$, is 
\begin{equation}
 \varepsilon \propto l^{-3}.
\end{equation}
For a displacement of one of the plates along the $z$-axis   $\delta
x^{3} \equiv \delta z = \delta l$. That
is why the force per unit area acting on a plate in terms of the energy
density  
$w = \varepsilon /l$ is 
\begin{equation}
F_z = -\frac{\delta \varepsilon }{\delta l} = 3w .
\end{equation}

At the same time as follows from (14)
\begin{equation}
F_z = \int_{\Sigma}<T^{33}>n^{3} da ,
\end{equation}
where $\Sigma$ is a surface of a plate, which has a unit area. 
It follows from the requirements $T^{\mu}_{\hspace{2 mm}\mu} = 0$ , 
$\partial_{\mu}T^{\mu \lambda} = 0$ and the symmetry of the problem that
for the case under 
consideration [6]
\begin{equation}
<T^{\mu \lambda}> = C(\frac{1}{4}g^{\mu \lambda} - \hat{x}^{\mu} 
\hat{x}^{\lambda})
\end{equation}
where $C$ is a constant, the metric $g^{\mu \lambda}$ has the signature 
$(-1, 1, 1, 1)$, 
and \linebreak $\hat{x}^{\mu} = (0,0,0,1)$ is a unit
vector 
along the $z$ axis. Thus, since $g^{00} = -1$ and \linebreak $g^{33} = 1$, 
\begin{equation} 
w \equiv <T^{00}> = -\frac{1}{4}C,
\end{equation}
and
\begin{equation}
T^{33} = (\frac{1}{4}g^{33} -1)C = -\frac{3}{4}C .
\end{equation}
 So, taking into account that 
$<T^{33>}$ is a constant we have from (19) 
\begin{equation}
F_z = <T^{33}> = -\frac{3}{4}C .
\end{equation}
Finally, from the comparison of this expression with (20) it follows that   
\begin{equation}
F_z = 3w ,
\end{equation}
which coincides with (17).

In the case of two perfectly conducting or two infinitely permeable
plates
the renormalized energy density $w$ is negative. So, as follows from (23), 
the force is attractive. In
the case of a perfectly conducting plate parallel to 
an infinitely permeable  plate $w$ is positive and therefore the force
is repulsive.


\begin{thebibliography}{20}
\bibitem{}
J. Sucher, private communication.
\bibitem{}
H. B. G. Casimir, Proc. K. Ned. Akad. Wet. {\bf 51} (1948) 793.
\bibitem{}
P. Debye, as reported in B. Chu, {\it Molecular Forces} (Wiley Interscience, 
        New York, 1967), p. 75.
\bibitem{}
A. Gonzalez,  Physica {\bf 131 A} (1985) 228.
\bibitem{}
P. Milonni, R. Cook and M. Goggin, Phys. Rev. A {\bf 38} 
       (1988) 1621.
\bibitem{}
L.S. Brown and G.J. Maclay, Phys. Rev. {\bf 184}  (1969) 1272. 
\bibitem{}
E. Hinds, Perturbative Cavity Quantum Electrodynamics, in {\it Cavity 
Quantum Electrodynamics}, edited by P. Berman (Academic Press, Boston,1994), 
p.30.
\bibitem{}
S. K. Lamoreaux, Phys. Rev. Lett. {\bf 78} (1997) 5.
\bibitem{}
G. Barton, J. Phys. A {\bf 14} (1981) 1009, ibid {\bf 15} 
(1982) 323. 
\bibitem{}
G. Barton,  Phys. Rep. {\bf 170} (1988) 1.
\bibitem{}
G. Barton in Ref. 7, p. 444.
\bibitem{}
V. Hushwater, Am. J. Phys. {\bf 65} (1997) 381.
\bibitem{}
S. Haroche, in {\it Fundamental Systems in Quantum Optics}, 
edited 
        by J. Dalibard, J. Raimond and J. Zinn-Justin (Elsevier Science, 
Amsterdam, 1992), p. 809. 
\bibitem{}
T. Boyer, Phys. Rev. A {\bf 9}  (1974) 2078.
\bibitem{}
J. Mathews and R. Walker, {\it Mathematical Methods of Physics}          
        (W. A. Benjamin, New York, 1965), p.343.
\end{thebibliography}
 \end{document}